\documentstyle[11pt,a4wide]{article}
\def\bkR{{\rm I\kern-.20em R}}
\begin{document}
   
\newcommand{\rot}{\vec{\nabla}\times}
\newcommand{\grad}{\vec{\nabla}}

\title{Lorentz gauge and Green's formula\\in\\classical electrodynamics}
\author{Nicolas Roy\\ \\Institut de F\'\i sica d'Altes Energies\\
  Universitat Aut\`onoma de Barcelona\\08193 Bellaterra (Barcelona),
  Catalonia, Spain}
\maketitle 

\section{Introduction}
When we are interesting in solving Maxwell's equations in
presence of charges, with given initial conditions for the electric
field $\vec{E}$ and the magnetic field $\vec{B}$, we can work with the
gauge field $A^{\mu}$ instead of the electric and magnetic fields.\\
 We thus have to choose a gauge.
We will see that the Lorentz gauge is a good gauge for such initial
value problems because Maxwell's equations for 
$A^{\mu}$ reduce to a set of uncoupled scalar wave equations that we can
integrate using the Green's method, and with restrictions on the possible
initial conditions for $A^{\mu}$. But we will also see that we have a
freedom in the choice of these initial conditions, and that we can always
choose them such that the influences of the charges and of the initial
excitation of the field are explicitly separated into the scalar and
the vector potential.
 
\section{Maxwell's equations in Lorentz Gauge}
If we call $\phi$ the time component $A^{o}$ of the gauge field, the
correspondence relations between the electric field $\vec{E}$, the
magnetic field $\vec{B}$ and the potentials $\vec{A}$ and $\phi$ are
\begin{eqnarray}\label{cores}
\begin{array}{l} 
    \displaystyle\vec{E}=-\frac{\partial}{\partial
      t}\vec{A}-\grad\phi\\
\\
    \displaystyle\vec{B}=\rot\vec{A}
\end{array}
\end{eqnarray}

Maxwell's equations for the gauge field $A^{\mu}$ read

\begin{equation}\label{max}
\partial_{\mu}\partial^{\mu}
A^{\nu}=J^{\nu}+\partial^{\nu}\partial_{\mu} A^{\mu}
\end{equation}
To work in Lorentz gauge, we have to impose the following {\em gauge
  fixing condition}  

\begin{equation}\label{fixing}
\partial_{\mu} A^{\mu}=0
\end{equation}
In this gauge, Maxwell's equations reduce to

\begin{equation}\label{maxlo}
\partial_{\mu}\partial^{\mu} A^{\nu}=J^{\nu}
\end{equation}
or explicitly, 
\begin{equation}
\label{eqrho}
\left\{
\begin{array}{l} 
        \displaystyle \Box\vec{A}=\vec{J}\\
\\
        \displaystyle \Box\phi=\rho
\end{array}
\right. \\
\end{equation}
where $\vec{J}$ is the current vector and $\rho$ the charge density.
We see in (\ref{eqrho}), that each component of $A^{\mu}$ is decoupled
from the others, and obeys a scalar wave equation of the form
\begin{equation}\label{eqnform}
\Delta\psi-\frac{1}{c^2}\frac{{\partial}^2}{\partial t^2}\psi =-4\pi
f(\vec{r},t)\\
\end{equation}

\section {Method of the Green's function}

To solve wave equations of the form (\ref{eqnform}), with given initial
conditions
$$
 \dot{\psi_o}(\vec{r})\equiv\left.\frac{\partial
    \psi(\vec{r},t)}{\partial t}\right|_{t=0} \ \mbox{ and }\ 
\psi_o(\vec{r})\equiv\psi(\vec{r},t=0) 
$$
we can use the Green's formula, which reads
\begin{equation}\label{greenformula}
4\pi\psi (\vec{r},t)=\frac{1}{c^{2}}\int_{\bkR^3}
d r_o^3\left(G\mid_{t_o=0}\dot{\psi_o}
-\frac{\partial}{\partial t_o}G\mid_{t_{0}=0}\psi_o\right)+
4\pi\int_{0}^{t} dt_o\int_{\bkR^3} d r_o^3f G
\end{equation}
where
\begin{equation}\label{greenfunction}
G\left(\vec{r},t\mid\vec{r_o},t_o\right)=\frac{1}{R}\delta\left(\frac{R}{c}-t+t_o\right)
\end{equation}
with $R=\left|\vec{r}-\vec{r_o}\right|$. The expression of $\psi$ is
thus composed of two terms, namely the {\em initial conditions term}
and the {\em source term}.

\section{Constraints on initial conditions}
Are the equations (\ref{max}) and (\ref{maxlo}) equivalent? Of course
not. In the general case, solutions of (\ref{maxlo}) are not solutions
of (\ref{max}) and actually not all possible initial conditions for
$\phi$ and $\vec{A}$ are allowed. 
Indeed, for the solutions of the reduced equation (\ref{maxlo}) to be solutions
of the general Maxwell's equation (\ref{max}), we have to choose
initial $A^\mu$ satisfying $\partial_{\mu} A^{\mu}=0$, and to ensure that
this gauge fixing condition is conserved in
time when integrating (\ref{maxlo}). 

\subsection{Stability of the gauge fixing condition}
Let's look at the evolution equation for $\Lambda=\partial_{\mu} A^{\mu}$. If we apply
$\partial_{\nu}$ to (\ref{maxlo}), we get
$$
\partial_{\nu}\partial_{\mu}\partial^{\mu} A^{\nu}=\partial_{\nu}J^{\nu}
$$
The RHS term vanishes because of charge conservation, and,
because derivative operators commute, we obtain
$$
\partial_{\mu}\partial^{\mu}\partial_{\nu} A^{\nu}=0
$$
The function $\Lambda$ thus obeys the equation 
$$
\Box\Lambda=0
$$
We can solve this equation with the method of Green's function, for
example, and we see that if at $t=0$ we take the following initial
conditions
$$
\left\{
\begin{array}{l} 
        \displaystyle \Lambda(\vec{r},0)=0 \\
        \displaystyle \dot{\Lambda}(\vec{r},0)=0
\end{array}
\right. \\
$$
then $\Lambda$ remain zero for all time.\\
\subsection{The constraints}
We thus have two constraints on the allowed initial conditions for $\phi$ and
$\vec{A}$. The first is obviously that our initial fields must
satisfy the gauge fixing condition which is explicitly 
$$
\Lambda=\dot{\phi}+\grad\vec{A}=0
$$
and the second is in fact nothing but Gauss' law. Indeed,
$\dot{\Lambda}$ is just
$$
\dot{\Lambda}=\ddot{\phi}+\grad \dot{\vec{A}}
$$
If we replace $\ddot{\phi}$, using (\ref{eqrho}), we obtain
\begin{eqnarray*}
\dot{\Lambda}&=&\Delta\phi+\grad \dot{\vec{A}}+\rho\\
             &=&\grad\left(\grad\phi+ \dot{\vec{A}}\right)+\rho
\end{eqnarray*}
If we remember the correspondence relations (\ref{cores}), the initial 
condition $\dot{\Lambda}=0$ reads
$$
\grad\vec{E}=\rho
$$
which is indeed Gauss' law.

\section{Convenient choice of the initial conditions}

Suppose we have a {\em physical initial situation}, i.e. we have
initial conditions for both electric and magnetic fields, which
we call $\vec{E}_o \mbox{ and }\vec{B}_o$ and which satisfy both
Gauss' law and $\grad\vec{B}=0$.\\
We then have to choose initial conditions for $\phi$ and $\vec{A}$
which provide the initial conditions $\vec{E}_o$ and $\vec{B}_o$ through
the relations (\ref{cores}), and which satisfy the gauge fixing
condition. We have a freedom for our choice, and we
will see that we can do it in a convenient way.\\

Indeed, because the gauge fixing constraint is only a restriction on
$\dot{\phi}$ and $\vec{A}$ and not on $\phi$ and $\dot{\vec{A}}$, we
can choose our initial scalar potential
$$
\phi_o =0
$$  
and {\em provide the whole initial excitation} of $\vec{E}$ from
$\dot{\vec{A}}$ alone, by setting
$$
\dot{\vec{A}}_o=-\vec{E}_o
$$ 
Now, we have to choose an initial $\vec{A}_o$ such that
$\vec{B}_o=\rot\vec{A}_o$. Once we have chosen this $\vec{A}_o$, then 
$\dot{\phi}_o$ is fixed by the gauge fixing condition via
$$
\dot{\phi}_o=-\grad\vec{A}_o
$$
But we have a freedom in the choice of $\vec{A}_o$ and, as we will
see, we can take it such that its divergence
vanishes\footnote{actually, this corresponds to taking the
  longitudinal part of $\vec{A}$ equal to zero.},
$\grad\vec{A}_o=0$, and thus $\dot{\phi}_o=0$. \\
For this, suppose we have $\vec{A}_o$ such that $\vec{B}_o=\rot
\vec{A}_o$. We see that we can add the gradient of a function,
namely $\grad f$, to $\vec{A}_o$ without changing the magnetic field.
For this new vector potential $\vec{A}_o+\grad f$, we have the new
 initial $\dot{\phi}_o$ 
$$
\dot{\phi}_o=-\grad\vec{A}_o-\Delta f
$$
If we want $\dot{\phi}_o$ to be zero, $f$ just has to satisfy
$$
\Delta f=-\grad\vec{A}_o
$$
and this equation has only one solution\footnote{if the field
  $\vec{A}_o$ vanishes at infinity}, given by inversing the Laplacian
operator, namely
$$
f\left(\vec{x}\right)=\frac{-1}{4\pi}\int dy^3\frac{-\grad\vec{A}_o\left(\vec{y}\right)}{\left|\vec{x}-\vec{y}\right|}
$$

We thus saw that given initial {\em physical}\footnote{statisfying
  Gauss' law and the divergence laws} $\vec{E}_o$ and $\vec{B}_o$, we can
choose the following initial conditions for the gauge field 

\begin{equation}\label{init}
\left\{
\begin{array}{l} 
        \displaystyle \phi_o(\vec{r}) =0 \\
        \displaystyle \dot{\phi}_o(\vec{r})=0\\
        \displaystyle \vec{A}_o(\vec{r}) \ \mbox{ such that }\  \vec{B}_o=\rot
                 \vec{A}_o \mbox{ and }\grad\vec{A}_o=0 \\
        \displaystyle \dot{\vec{A}}_o(\vec{r})=-\vec{E}_o
\end{array}
\right. 
\end{equation}

\section{Final expressions}
With the initial conditions previously found, we are now able to
express the solutions of the Maxwell's equations via the Green's formula.
For the vector potential, the source term is the current $\vec{J}$ and
the expression is

\begin{equation}\label{aformula}
4\pi\vec{A} (\vec{r},t)=\frac{-1}{c^{2}}\int
d r_o^3\left(G\mid_{t_o=0}\vec{E}_o
+\frac{\partial}{\partial t_o}G\mid_{t_{0}=0}\vec{A}_o\right)+
\int_{0}^{t} dt_o\int d r_o^3 \ G\ \vec{J}
\end{equation}
which doesn't depend on the charge distribution. 
For the scalar potential, the source term is the current $\rho$ and
its expression is

\begin{equation}\label{phiformula}
4\pi\phi (\vec{r},t)=\int_0^t dt_o\int d r_o^3 \ G\ \rho
\end{equation}
which depends only on the charge distribution.

\section{Conclusion}

The Lorentz gauge is convenient for treating initial conditions problems,
because in this gauge, Maxwell's equations become wave equations,
which we can solve using the method of Green's function. \\
In the last section we saw, that given initial {\em
  physical}\footnote{statisfying Gauss' law and the divergence laws} $\vec{E}$ and
$\vec{B}$, we can choose convenient initial conditions (\ref{init})
for the gauge field, and with this choice, the contributions of the
charges and the initial excitation of $\vec{E}$ and $\vec{B}$ are
well separated because 
\begin{itemize}
\item the solution for the scalar potential $\phi$, is totally
independent of the initial excitations of $\vec{E}$ and $\vec{B}$, and
depends only on the charge distribution $\rho$.
\item the solution for the vector potential $\vec{A}$ depends on the
  current distribution $\vec{J}$, and on the initial excitation of the
  electromagnetic field, and not on the charge distribution.
\end{itemize}
Moreover, if we only consider fixed charge distributions, then the
scalar potential $\phi$ depends only on the charges, and the vector
potential $\vec{A}$ depends only on the initial excitation.

\end{document}